\documentclass[epj]{svjour}
\usepackage{graphics}
\usepackage{amssymb}
\usepackage{amsmath}
\usepackage[dvips]{epsfig}
\newcommand{\fmn}[2]{\mbox{${\textstyle \frac{#1}{#2}}$}}
\begin{document}
\title{Exchange Reactions with Dick Dalitz}
\author{D.H.~Davis}
\institute{Department of Physics \& Astronomy,
University College London, London WC1E 6BT, UK}
\date{Received: \today / Revised version:}
\abstract{An account is given of the close collaboration of Dick
Dalitz with the European $K^-$ Collaboration over many decades in
many aspects of hypernuclear physics. In particular, emphasis is
given to the topics of double hypernuclei and the discovery and
resolution of $p$--wave $\Lambda$ strangeness exchange states.
 } \maketitle

\section{Introduction}

This is a personal account of a career of 50 years which strongly
overlapped with two of the many of Dick Dalitz's interests,
hypernuclei and the low energy $\bar{K}N$ system.

As a new graduate student, I first met Dick about 50 years ago
when he gave a seminar on the low energy $\bar{K}N$ system, a
preview of the subsequent Dalitz and Tuan paper~\cite{ref1}.

My first publication~\cite{ref2} concerned the observation of a
$\Sigma^+$ hyperon decaying into a proton with an associated
Dalitz pair~\cite{ref3}, a mere verification of what was already
well known; one decay mode of the $\Sigma^+$ was to $\pi^0p$.

In 1961 I joined Levi Setti's emulsion group in Chicago, the hub
of hypernuclear physics at that time, and got to know Dick very
well.  While there, we made the first direct experimental estimate
of the $\Lambda$ nuclear potential well-depth, $D_{\Lambda}$, from
the decays of heavy spallation hypernuclei~\cite{ref4}.  Our
value, of a little less than 30\,MeV, agreed well with Dick's
value of $26.5\pm2.5$\,MeV~\cite{ref5} obtained by extrapolating
from the then known binding energy of $_{\,\Lambda}^{13}$C.
However, it should be remarked that at this time various
theoretical estimates using $\Lambda N$ potentials derived from
the $B_{\Lambda}$ values of the $s$--shell hypernuclei all gave
values nearer to 60\,MeV. Also at this time was the determination
of the spin of $_{\Lambda}^{\,8}$Li~\cite{ref6,ref7}.

I returned to London to rejoin the European $K^-$ Emulsion
Collaboration and Dick came to England shortly afterwards, first
to Cambridge and then permanently to Oxford to join up again with
his old mentor, Rudolf Peierls.

There began a long, symbiotic, and fruitful collaboration between
Dick and us. We benefited greatly from his advice, encouragement
and theoretical input, and he knew of our results well before
publication. He was invited and came to many of our meetings, not
only those in London, but sometimes to those in Brussels and CERN
as well.

In the sixties and early seventies we worked on such topics as the
$B_{\Lambda}$ values of $p$--shell hypernuclei~\cite{ref8}, spins
of $_{\Lambda}^{\,8}$Li~\cite{ref9} and
$_{\,\Lambda}^{12}$B~\cite{ref10}, $\pi^+$ decays of
hypernuclei~\cite{ref11}, charge symmetry breaking in the
$_{\Lambda}^{\,4}$He, $_{\Lambda}^{\,4}$H mirror
doublet~\cite{ref12} and many others, always in close
collaboration with Dick.

However, I would like to concentrate on two long-running sagas in
which Dick played a pivotal role, those of double hypernuclei and
the proton--emitting $p$-wave hypernuclear states.

\section{Double Hypernuclei versus the H}

The first double hypernucleus, an example of $_{\Lambda\Lambda}^{\
10}$Be, was found by our collaborators in Warsaw in
1962~\cite{ref13} and Dick was among the first to glean properties
of the $\Lambda\Lambda$ interaction from it~\cite{ref14}. A second
event, an example of $_{\Lambda\Lambda}^{\ \ 6}$He, was reported
by Prowse in 1966~\cite{ref15} and then, for a long time, nothing.
Prowse died in 1971 and Pniewski began to express doubts about the
authenticity of the Prowse event.  In 1977, Jaffe predicted the
existence of the $H$ particle, a deeply bound six quark system
$(u,u,d,d,s,s)$, the quark content of two $\Lambda$ hyperons.  I
well remember questioning Jan de Swart at the Jab{\l}onna
Conference in 1979, that was it not difficult to reconcile the
existence of double hypernuclei with that of the $H$ particle? His
reply, which was echoed in many subsequent reviews by proponents
of the $H$, was that there were only two events reported in
emulsions and perhaps they had been wrongly interpreted.

Dick asked me at the end of the eighties if we could possibly
remeasure the event in order to allay these doubts.  I told him
that this was not possible as the emulsion pellicle containing the
event no longer existed.  However, I did have in my possession
copies of unpublished photomicrographs taken of the event by
P.\,Fowler \emph{et al.}\ in Bristol some 25 years earlier.

It should be noted that the three vertices of the event, the
$\Xi^-$ capture and the double and single hypernuclear decays were
all contained in a cube of side length 3 microns.  Moreover,
nuclear emulsion is approximately one half silver bromide crystals
by volume and, as a consequence, a pellicle shrinks in height by a
factor a little more than two on processing.  In order to achieve
greater resolution for photography, the emulsion was swelled by
soaking it in a saturated sugar solution made from Tate and Lyle's
golden syrup, since water has the wrong optical properties, to 3.5
times its processed thickness. Unfortunately for Dick's request,
the sugar ultimately crystallised and the emulsion was destroyed.
The photomicrographs and an independent analysis of the event made
during its photographing were published in the Royal Society paper
of 1989~\cite{ref16}.

Every attempt was made to elicit further details of the Prowse
event but without success.  However, one significant fact did
emerge which persuaded me that Pniewski was right to doubt the
event.  Although Prowse had moved to Wyoming and published the
event from there, his emulsion work had remained in UCLA and where
at that time Ticho, Schlein and Slater, all erstwhile hypernuclear
emulsion physicists, were based.  All were contacted, but none had
any recollection of having seen the event!

There now exists the tightly constrained $_{\Lambda\Lambda}^{\ \
6}$He~\cite{ref17} with a $\Delta B_{\Lambda\Lambda}\approx
1.0\,$MeV, completely at odds with the Prowse event, and for that
matter with the original $_{\Lambda\Lambda}^{\ 10}$Be event also.
For the $_{\Lambda\Lambda}^{\ 10}$Be event, as was stated in a
footnote to a table in the paper of 1963, `It should be noted that
the value of $B_{\Lambda\Lambda}(_{\Lambda\Lambda}Z)$, and hence
$\Delta B_{\Lambda\Lambda}$, may be overestimated if the ordinary
hyperfragment is produced in an excited state', and should the
decay have occurred to a now known state of $_{\Lambda}^{\,9}$Be
at $\sim3.0\,$MeV~\cite{ref18}, the $\Delta B_{\Lambda\Lambda}$
would become $\sim 1.3\,$MeV, well compatible with the Nagara
event~\cite{ref17}. No such escape clause exists for the Prowse
$_{\Lambda\Lambda}^{\ \ 6}$He event~\cite{ref15}.

\centerline{And still there is no $H$!}

\section{Strangeness--Exchange States}

Around the late sixties we were attempting to determine the
$B_{\Lambda}$ values of $p$--shell hypernuclei.  There were many
possible examples of $_{\,\Lambda}^{11}$B decays to
$\pi^-\:{}_{\,\Lambda}^{11}$C but see Table~\ref{table1}.

\begin{table}[htb]
\begin{center}
\caption{Competing two--body decay characteristics \label{table1}}
\begin{tabular}{cccc}
\hline
&pion range&recoil range& recoil\\
&($\mu$m)&($\mu$m)&\\
\hline
$_{\,\Lambda}^{11}\textrm{B}\to\pi^-\:{}_{\,\Lambda}^{11}$C&
20700&1.0&$\beta^+$\\
$_{\,\Lambda}^{10}\textrm{Be}\to\pi^-\:{}_{\,\Lambda}^{10}$B&19800&1.1&
stable\\
$_{\Lambda}^{\,7}\textrm{Li}\to\pi^-\:{}_{\Lambda}^{\,7}$Be&21600&1.8&E.C.\\
\hline
\end{tabular}
\end{center}
\end{table}

With range straggling of the pions of about 3\%, the three
hypernuclei in Table~\ref{table1} cannot be separated from decay
kinematics alone. It was noted that many of the
$_{\,\Lambda}^{11}\textrm{B}$ candidates had $K^-$ at rest
production topologies of hyperfragment + $\pi$ + one baryonic
track. The azimuthal and dip angles of all three tracks were
measured, as were the ranges of the hyperfragment and assumed
proton (the pion invariably left the emulsion before stopping).

Assuming the $K^-$ capture is on a light nucleus of the emulsion,
in this case $^{12}$C, the proposed reaction is
\begin{equation}
\label{eq1} K^- + ^{12}\textrm{C}\rightarrow \pi^- + p +
_{\,\Lambda}^{11}\textrm{B} + Q.
\end{equation}
The range of the proton determines both $T_p$ and $p_p$. With the
$Q$ value known, an iterative procedure was used to determine the
value of the pion's energy.  Starting with a value of $T_{HF} =
0$, this gave $T(1)_{\pi}$, hence $\vec{p}(1)_{\pi}$,
$\vec{p}(1)_{\pi} + \vec{p}_p \rightarrow \vec{p}(1)_{HF}
\rightarrow T(1)_{HF} \rightarrow T(2)_{\pi} \rightarrow
\vec{p}(2)_{\pi}$, $\vec{p}(2)_{\pi} + \vec{p}_p\rightarrow
T(2)_{HF}$, and so on. The procedure rapidly converges; after two
iterations the pion kinetic energy is stable to 5 keV. The
compatibility with reaction (\ref{eq1}) was then tested by
comparing the measured and computed ranges and directions of the
hypernucleus. The compiled data revealed a sharp spike, of the
order 1\,MeV wide, in the pion spectrum.  Such a spike suggests
that many of the events proceeded \emph{via} a two-body reaction
\begin{eqnarray}
\nonumber K^- +
^{12\!}\textrm{C}\rightarrow\pi^-+_{\,\Lambda}^{12}\textrm{C}^*&&\\
&&\hspace{-5mm}\hookrightarrow p +_{\,\Lambda}^{11}\textrm{B}
\label{eq2}
\end{eqnarray}

This was written up and sent to Nuclear Physics but the referee
demurred --- `Not every spike implies a resonance.' It was
resubmitted with the inclusion of phase space and impulse model
curves. The spike remained prominent, but still the referee was
not happy, and so it went to a second referee.  The second referee
was more forthcoming, `first observation of a highly excited state
of a hypernucleus', and so it was published~\cite{ref19}.  I saw
Dick shortly afterwards and thanked him for overruling the first
referee.  He was taken aback and said that Nuclear Physics had no
business to divulge the referee's name.  I had to assure him that
they had not.  Remember, this was well before the coming of the
Word, the Microsoft Word that is.  Referees wrote reports which
were typed by a secretary.  However, as anyone who has received a
typewritten document from Dick will know, he was never satisfied
and always modified the text in his own neat but unmistakable
handwriting, and so it was here.

We published a further paper on the subject with some more
statistics but then left the field to go hunting for charm and
beauty particles.  The subject was taken up by the counter
$(K,\pi)$ and $(\pi,K)$ spectroscopy groups at CERN, BNL and later
KEK.  I was brought back to the subject by Dick, who else,
approaching me at the 1982 Heidelberg Conference.  He had a
problem. Was the $\Lambda$ in $_{\,\Lambda}^{12}\textrm{C}^*$
bound or unbound? Theoreticians worry about such things. The
Brookhaven group had two values for $B_{\Lambda}$ in
$_{\,\Lambda}^{12}\textrm{C}^*$, one measured with the pion going
off in the forward direction as everyone else had done, the other
with the pion making a large, $15^{\circ}$ angle to the kaon
direction.
\begin{eqnarray*}
\theta_{K\pi} = \phantom{1}0^{\circ}\ \  &\Rightarrow&\ \
B_{\Lambda} = +0.033\pm
0.180\,\textrm{MeV},\\
\theta_{K\pi} = 15^{\circ}\ \  &\Rightarrow&\ \  B_{\Lambda} =
-0.027\pm 0.160\,\textrm{MeV}.
\end{eqnarray*}

\begin{figure}[htb]
\begin{center}
\resizebox{0.45\textwidth}{!}{%
 \includegraphics{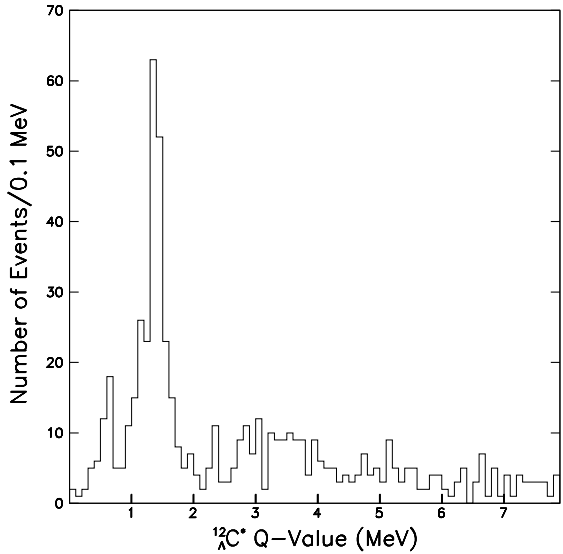}}
\end{center}
\caption {Distribution of the $Q$-values in the
$_{\,\Lambda}^{12}\textrm{C}^*\to p +_{\,\Lambda}^{11}\textrm{B}$
decay, treating all events as occurring on $^{12}$C.\label{fig1}}
\end{figure}

He remarked that he had also checked our two papers and in both
the $B_{\Lambda}$ values were positive, but they did not agree
with one another! This was a problem.  Dave Tovee and I went back
to our old results and discovered the cause of the discrepancy;
the input mass values necessary to our analysis, especially that
of the kaon, had changed in the Particle Data Group's listings in
the early seventies by considerably more than the stated errors.
We reanalysed all of our old data, including a large sample of
events obtained using the resonance peak to identify
$_{\,\Lambda}^{11}\textrm{B}$ events in order to study their
non-mesonic decays~\cite{ref20} and the results of this analysis
were presented at the 1985 Brookhaven Conference~\cite{ref21}. The
$Q$ value distribution in the decay
$_{\,\Lambda}^{12}\textrm{C}^*\to p +_{\,\Lambda}^{11}\textrm{B}$
is given in Fig.~\ref{fig1}~\cite{ref22}.

A good fit to the spectrum below 2\,MeV, the events there comprise
an essentially pure sample of $_{\,\Lambda}^{12}\textrm{C}^*$,
requires at least three Breit--Wigner distributions.  One stands
alone, but two more are needed, both centred around
$Q\approx1.4\,$MeV, a very narrow one and a wider one. What do we
expect?  The states are formed by the addition of a $P\fmn{3}{2}$
neutron hole and either a $P\fmn{1}{2}$ or a $P\fmn{3}{2}$
$\Lambda$ hyperon.
\begin{eqnarray*}
\bar{n}(P\fmn{3}{2}) + \Lambda(P\fmn{1}{2})&\Rightarrow& 1^+,2^+,\\
\bar{n}(P\fmn{3}{2}) + \Lambda(P\fmn{3}{2})&\Rightarrow&
0^+,1^+,2^+,3^+.
\end{eqnarray*}

The low energy $\bar{K}N$ interaction is $s$--wave, which contains
no spin flip, and so we expect only the $0^+$ and the two $2^+$
states to be produced. Since the $_{\,\Lambda}^{11}\textrm{B}$
ground state has spin $\fmn{5}{2}^+$, while the proton from the
two $2^+$ states may be emitted in the $s$--wave, that from the
$0^+$ state has to be $d$--wave.  It is thus natural to assign the
observed narrow state to be the $0^+$, since the angular momentum
barrier for such a small energy release, $\sim 1\,$MeV, will
strongly inhibit its decay. With this assignment we construct the
level scheme given in Table~\ref{table2}.

\begin{table}[htb]
\begin{center}
\caption{Level scheme for the observed
$_{\,\Lambda}^{12}\textrm{C}^*$ states \label{table2}}
\begin{tabular}{cccc}
\hline
&No of&$\Gamma$&$B_{\Lambda}$\\
&events&(keV)&(MeV)\\
\hline%
$0^+$&  64 &     $<100$ & $+0.14\pm0.05$\\
$2^+$& 193 & $\sim 600$ & $+0.20\pm0.05$\\
$2^+$&  48 & $\sim 150$ & $+0.95\pm0.05$.\\
\hline
\end{tabular}
\end{center}
\end{table}

To conclude, three states are expected, and three are found.  The
real widths of the $2^+$ states have been determined, whereas it
is only possible to give an upper limit to the width of the $0^+$
state; there are limits even to the resolution of the emulsion
technique! The $B_{\Lambda}$ difference between the two $2^+$
states places a limit on the $\Lambda N$ spin--orbit interaction
and the relative production rates following $K^-$ capture at rest
on $^{12}$C are given. Finally, in answer to Dick's original
question, the $B_{\Lambda}$ values have been determined and in ALL
states the $\Lambda$ hyperon is bound.

The remaining problem with this analysis; why did the $0^+$ state
not decay quickly by $s$-wave proton emission to the expected
$\fmn{1}{2}^+$ excited state of $_{\,\Lambda}^{11}\textrm{B}$ was
solved recently when it was found that this state was
energetically out of reach~\cite{ref23}.

\section{Conclusions}

In conclusion, it should be emphasised that neither of these two
investigations would have been undertaken without Dick's
encyclopaedic knowledge of past results and his nagging
persistence to obtain solutions to puzzling situations.  The
relevant information would otherwise have mouldered away on my
bookshelves, forgotten.

I greatly miss his friendship and his phone calls, usually at home
late on a weekend evening, asking for details of things recorded
in his notebooks of which I had presumably told him some twenty or
thirty years before.  The whole hypernuclear community will sorely
miss him too.

I would like to thank the organisers for giving me the opportunity
to speak of Dick, for a very enjoyable conference and for
contributing generously towards my expenses.

%
%%%%%%%%%%%%%%%%%%%%%%%%%%%%%%%%%%%%%%%%%%%%%%%%%%%%%%%%%%%%%%%%%%%%%%%%%%%
%

\end{document}